# Interplay between nonlinear spectral shift and nonlinear damping of spin waves in ultrathin YIG waveguides

S. R. Lake[1], B. Divinskiy[2*], G. Schmidt[1,3], S. O. Demokritov[2], and V. E. Demidov[2]

[1]*Institut für Physik, Martin-Luther-Universität Halle-Wittenberg, 06120 Halle, Germany*

[2]*Institute of Applied Physics, University of Muenster, 48149 Muenster, Germany*

[3]*Interdisziplinäres Zentrum für Materialwissenschaften, Martin-Luther-Universität Halle-Wittenberg, 06120 Halle, Germany*

We use the phase-resolved imaging to directly study the nonlinear modification of the wavelength of spin waves propagating in 100-nm thick, in-plane magnetized YIG waveguides. We show that, by using moderate microwave powers, one can realize spin waves with large amplitudes corresponding to precession angles in excess of 10 degrees and nonlinear wavelength variation of up to 18% in this system. We also find that, at large precession angles, the propagation of spin waves is strongly affected by the onset of nonlinear damping, which results in a strong spatial dependence of the wavelength. This effect leads to a spatially-dependent controllability of the wavelength by the microwave power. Furthermore, it leads to the saturation of nonlinear spectral shift's effects several micrometers away from the excitation point. These findings are important for the development of nonlinear, integrated spin-wave signal-processing devices and can be used to optimize their characteristics.

*Corresponding author, e-mail: b_divi01@uni-muenster.de

# I. INTRODUCTION

Nonlinear phenomena accompanying propagation of spin waves in magnetic films have been known for many decades to enable implementing a large variety of advanced signal-processing devices [1,2]. One fundamental nonlinear phenomenon that is frequently applied is the nonlinear transformation of the spin wave's dispersion spectrum that occurs with increasing spin wave intensity. This phenomenon can be regarded as a frequency shift of the dispersion spectrum of the spin waves. This shift results from the decrease in the magnitude of the static magnetization caused by the increase in the magnetization precession angle [1,2]. Because of this frequency shift, the wavelength of a spin wave at a given frequency becomes dependent on its amplitude, which allows one to implement, for example, nonlinear magnonic phase shifters [3,4], interferometers [5], couplers [6,7], and switches [8]. Additionally, these effects hold promise for implementation of spin-wave logic devices [9-11] and neuromorphic computing with spin waves [12,13].

The vast majority of previous work on nonlinear spin waves utilized millimeter-scale magnetic structures fabricated from low-loss magnetic insulator – yttrium iron garnet (YIG) [14]. In recent years, the miniaturization of spin-wave devices down to micrometer and sub-micrometer dimensions has become the main trend and challenge in the field of magnonics. Such miniaturization has been greatly advanced by the advent of high-quality ultrathin YIG films [15-17] that can be structured on the nanometer scale [18-20]. This breakthrough not only provides the possibility to implement low-loss propagation of spin waves in the linear regime, but also brings novel opportunities to investigate dynamic nonlinear phenomena in ultrathin films and their utilization in signal-processing applications. In particular, the strong quantization of the spectrum of magnetic excitations in microscopic YIG structures can substantially suppress the spin-wave scattering effects and help achieve very large amplitudes of magnetic dynamics [21] that have not been observed on the macroscopic scale.

Although nonlinear spin-wave dynamics in microscopic YIG structures remain largely unexplored, it has already been demonstrated experimentally that the nonlinear shift of the dispersion spectrum of propagating spin waves can be used, for example, to control switching in coupled spin-wave waveguides via intensity [7] and to indirectly excite propagating spin waves [22]. It has also been theorized to be the underlying physical phenomenon for the realization of nonlinear nano-ring resonators [23] and nanoscale neural networks [13]. However, until now it has remained unclear how large nonlinear shift is practically achievable and which factors can limit this effect in real devices.

In this work, we study the propagation of intense spin waves in microscopic ultrathin-YIG waveguides using a wide range of power for the excitation signal. By using high-resolution, phase-sensitive, magneto-optical detection, we directly measure the spin wave's spatial dependencies of intensity and wavelength. Our experimental findings indicate that the nonlinear spectral shift is strongly affected by the onset of the nonlinear magnetic damping: at large amplitudes, spin waves start to exhibit strongly enhanced spatial attenuation leading to a spatially-dependent magnitude of the nonlinear spectral shift. As a result, the wavelength (wavevector) of intense spin waves varies strongly along the propagation path. While the wavelength exhibits good controllability by the applied microwave power at the beginning of the propagation path, this controllability is almost completely lost several micrometers away from the excitation point. Our findings provide insight into nonlinear propagation of spin waves in microscopic waveguides and are of decisive importance for the development of efficient magnonic devices utilizing the effect of the nonlinear spectral shift.

## II. EXPERIMENT

Figure 1(a) shows the schematics of our experiment. We study propagation of spin waves in a 2-μm wide spin-wave waveguide patterned from a 100-nm thick YIG film. The YIG film is characterized by a saturation magnetization of $4\pi M_S = 1.75$ kG and a Gilbert damping constant $\alpha = 4\times10^{-4}$, as determined from ferromagnetic resonance measurements. The spin waves are excited by using a 500-nm wide and 150-nm thick inductive Au antenna carrying microwave current with a frequency $f$ and a power $P$. The YIG waveguide is magnetized to saturation by a static magnetic field, $H_0$=1000 Oe, which is applied in-plane along the Au antenna.

For patterning, a double-layer PMMA resist is deposited on a <111> oriented GGG substrate. The resist is exposed by e-beam lithography using a RAITH Pioneer system and developed in isopropanol. Subsequently 110 nm of YIG are deposited at room temperature by pulsed laser deposition using a recipe by Hauser et al. [17]. After lift-off in acetone, the sample is annealed in a pure oxygen atmosphere. Wet etching in Phosphoric acid is used to remove 10 nm of YIG to smoothen the edges of the remaining structures. The process is completed by patterning a microstrip antenna (10 nm Ti, 150 nm Au) on top of the structures using electron beam lithography, e-beam evaporation, and lift-off.

We study the propagation of spin waves in the YIG waveguide with spatial and phase resolution by using micro-focus Brillouin light scattering (BLS) spectroscopy [24]. We focus the probing laser light with a wavelength of 473 nm and a power of 0.25 mW into a diffraction-limited spot on the sample surface [Fig. 1(a)] and analyze the modulation of the probing light due to its interaction with the magnetization dynamics. The intensity of this modulation, or BLS intensity, is proportional to the intensity of spin waves at the position of the probing spot. This allows us to record two-dimensional maps of the spin-wave intensity by rastering the spot over the sample surface. Additionally, we use the interference of the modulated light with the reference light to measure the spatial maps of $\cos(\varphi)$, where $\varphi$ is the phase difference between the spin wave and the microwave signal applied to the antenna. Analysis of the phase maps provides information about the wavelength of spin waves at a given frequency $f$ and allows us to directly address the effect of the nonlinear spectral shift.

First, we characterize the expected effect by using micromagnetic simulations (Figs. 1(b) and 1(c)). We calculate amplitude-dependent dispersion curves using the simulation software MuMax3 [25] and the approach developed in Ref. [26]. We consider a 2-µm wide and 100-nm thick waveguide with the length $L$=20 µm discretized into 10 nm × 10 nm × 10 nm cells with periodic boundary conditions at the ends. The standard for YIG exchange constant of 3.66×10$^{-7}$ erg/cm is used. The Gilbert damping parameter is set to an artificially small value 10$^{-12}$ to fix the amplitude of the magnetization dynamics at the chosen level. We excite magnetization dynamics by initially deflecting magnetic moments from their equilibrium orientation by an angle $\theta$. The deflection is spatially periodic with the period $L/n$ (where $n$ is an integer), which defines the wavelength of the excited wave. By analyzing the free dynamics of magnetization, we determine the frequency corresponding to the given spatial period and obtain the frequency vs wavenumber relations (solid curves in Fig. 1(b)) for a given spin-wave amplitude characterized by the precession angle $\theta$.

As seen from the data of Fig. 1(b), the increase in the angle $\theta$ from 0.1 to 15° leads to a noticeable shift of the dispersion curve down toward smaller frequencies and a slight overall decrease in its slope. Both effects are consistent with the changes expected for a decrease in the static component of magnetization, $M_{ST}$, with the increase of the amplitude of the magnetization precession: $M_{ST} = \sqrt{(M_S^2 - m^2)} \approx M_S - \frac{1}{2}M_S\sin^2(\theta)$, where $M_S$ is the saturation magnetization, $m$ is the amplitude of the dynamic magnetization, and $\theta$ is the mean precession angle. Because $M_{ST}$ enters the expression for the frequency of ferromagnetic resonance: $f_{FMR} = \gamma\sqrt{H_0(H_0 + 4\pi M_{ST})}$, its decrease causes the overall decrease in the frequency of spin waves. Additionally, the decrease of $M_{ST}$ is known to lead to a decrease in the group velocity of spin waves [2], which is consistent with the overall decrease of the slope of the dispersion curve in Fig. 1(b). The nonlinear transformation of the spectrum can be quantitatively characterized by the variation of the wavenumber d$k$ due to the increase in $\theta$ from 0.1 to 15° (dashed curve in Fig. 1(b)). These data indicate that the dispersion's shift becomes significantly stronger as wavelength decreases. As can be seen from the data of Fig. 1(c), d$k$ varies approximately linearly with $\sin^2(\theta)$, which in turn is proportional to the spin-wave intensity. This linear dependence makes it straightforward to calibrate the precession angle achieved in the experiment. We also emphasize that the dispersion curve calculated for $\theta = 0.1°$ coincides well with that measured by BLS at low excitation power $P = 0.1$ mW. This provides clear proof of the validity of the results obtained from the micromagnetic simulations.

## III. RESULTS AND DISCUSSION

To experimentally address the effect of the nonlinear spectral shift, we perform phase-resolved BLS measurements at a fixed frequency of the excitation signal $f$ and analyze the variation of the wavelength of spin waves with the increase in the power $P$. Figure 2(a) shows representative spin-wave phase maps recorded at $f = 4.8$ GHz and $P = 0.1$ and 3 mW, whereas Fig. 2(b) shows the spatial dependence of the wavelength of spin waves, $\lambda$, obtained from the Fourier analysis of these maps. As seen from these data, at $P = 0.1$ mW, the spin waves exhibit a well-defined constant wavelength $\lambda = 1.38$ μm, which is in good agreement with the dispersion curve calculated for the small precession angle $\theta = 0.1°$ (Fig. 1(b)). At $P = 3$ mW, the situation changes drastically. First, the wavelength is reduced, explained by the effect of the spectral shift. Second, the wavelength strongly changes in space exhibiting the maximum reduction of about 18% close to the antenna, which becomes as small as 3% at the propagation distance of 18 μm. This spatial variation could be associated with the decrease of the intensity of the spin wave along the propagation path due to the damping. However, taking into account the linear dependence in Fig. 1(c), this would require that the intensity decreases by more than a factor of five over the propagation interval $x = 2 – 18$ μm, which is inconsistent with the small damping in the YIG film.

We quantify the spatial decay of spin waves in the waveguide by analyzing the spin-wave intensity maps (Fig. 3(a)). The direct comparison of the maps recorded at $P = 0.1$ mW and 3 mW clearly shows that the spatial decay becomes significantly stronger, as the power is increased. In Fig. 3(b), we show in the log-linear coordinates the spatial dependence of the BLS intensity integrated over the waveguide width. At small power, $P = 0.1$ mW, this dependence is exponential and is characterized by the decay length $\xi = 29$ μm, where $\xi$ is the distance, at which the amplitude decreases by a factor of e. This value agrees reasonably well with $\xi = 34$ μm obtained from micromagnetic simulations using experimentally determined $\alpha = 4\times10^{-4}$. In contrast to these simple behaviors, at large powers, the spatial decay is not described by a single exponential and strongly exceeds that observed in the linear propagation regime. This fact explains the strong spatial variation of the wavelength in Fig. 2(b).

To characterize the modification of the decay with the increase of $P$ in detail, we plot in Fig. 4(a) the power dependence of the BLS intensity recorded at $x = 0$ and 20 μm. We note that, in the linear propagation regime, these dependences are expected to grow linearly with power. As seen from the data of Fig. 4(a), at $x = 0$, the BLS intensity remains proportional to $P$ at $P < 1$ mW and then starts to saturate. At approximately the same power, the intensity recorded at $x = 20$ μm exhibits a drop reflecting an increase in the spatial decay. This indicates an onset at the threshold power $P = 1$ mW of the so-called nonlinear damping [26-32], which is associated with the energy transfer from the coherent propagating spin wave into other incoherent short-wavelength spin-wave modes. These short-wavelength spin-wave modes do not contribute to the BLS intensity and cannot be directly accessed in the experiment.

The contribution of this phenomenon relative to the normal linear damping can be estimated based on the data presented in Fig. 4(b). The figure shows the decrease of the spin wave intensity $I$ over a propagation length of 1 µm expressed by an attenuation factor $I(x)/I(x+1$ µm). This factor is plotted over the used range of microwave power for positions $x = 0$ and 20 um. At large distances from the antenna ($x = 20$ µm), the attenuation factor is about 1.08 and remains approximately constant within the entire studied interval of $P$ (point-down triangles in Fig. 4(b)). This value is consistent with that expected for the effects of the linear damping for $\alpha = 4\times10^{-4}$. We note that it does not increase at large $P$, because even for $P = 3$ mW, the strong nonlinear decay at the initial propagation stage has already severely diminished the intensity of spin waves at $x = 20$ µm. This initial strong decay is well characterized by the attenuation factor measured at $x = 0$ (point-up triangles in Fig. 4(b)). In the linear propagation regime ($P < 1$ mW), the attenuation factor remains constant and coincides with the observed value at $x = 20$ µm. However, at $P > 1$ mW, it increases significantly and reaches the value of 1.93 at $P = 3$ mW. Comparing this value with 1.08, one can conclude that the nonlinear energy transfer into short-wavelength spin-wave modes causes the increase of the effective damping of the coherent wave by nearly a factor of two.

The results presented above suggest that, due to the effects of the nonlinear damping, the efficient controllability of the wavelength (wavenumber) by the microwave power can only be achieved near the excitation point, while at large propagation distances this controllability becomes relatively poor. This is evidenced by the data in Fig. 5, which shows how the change in wavenumber, d$k$, depends on the applied power for various distances from the antenna. Close to the antenna, d$k$ changes linearly with the microwave power reaching the value of 0.9 µm$^{-1}$ at $P = 3$ mW (point-up triangles in Fig. 5). In contrast, at a distance $x = 8$ µm (circles in Fig. 5), d$k$ saturates to the value of about 0.4 µm$^{-1}$ and remains nearly constant for $P > 1.5$ mW. Finally, at $x = 18$ µm (point-down triangles), the maximum achieved shift never exceeds 0.1 µm$^{-1}$ within the entire range of $P$.

It is instructive to discuss approaches to suppress the detrimental effects of the nonlinear damping. The dominating mechanism of the nonlinear damping is the energy transfer from a coherent spin wave into incoherent modes possessing the same frequency. This process can be controlled by varying the geometrical parameters of the waveguide, which allows one to modify the spin-wave dispersion spectrum using the effects of spin-wave quantization and avoid the detrimental spectral degeneracy (see, e.g., Refs. 32, 33). In particular, this can be achieved by reducing the width of the waveguide below a certain critical value. For the 100-nm thick YIG film used in our study, we estimate the critical width of about 200 nm. Reduction of the width below this value can allow a significant suppression of the nonlinear damping within the addressed range of the wavelength of spin waves. However, this reduction is also expected to result in an increase of the linear damping due to the increasing contribution of the spin-wave scattering caused by the roughness of the waveguide edges. We note that, due to the competing effects of the dipolar and the exchange interaction, the critical width weakly depends on the thickness of the YIG film. It remains nearly unchanged within the practically important range of thicknesses 50-200 nm. Additionally, since the nonlinear damping is caused by the parametric interactions of spin-wave modes, its threshold can be increased by increasing the linear damping. However, this approach is impractical, since it leads to a faster decay of spin waves. Finally, the threshold power of the nonlinear damping can be increased by reducing the ellipticity of the magnetization precession, which can be achieved by using magnetic films with perpendicular magnetic anisotropy [26,34].

We now turn to the estimation of characteristic precession angles $\theta$ in our experiments. We base our analysis on the comparison of the experimental dependence d$k(P)$ measured close to the antenna (point-up triangles in Fig. 5) with the dependence d$k(\sin^2\theta)$ obtained from micromagnetic simulations (red curve in Fig. 1(c)). Because both of these relationships are linear with their respective variable, one can conclude that $\sin^2\theta$ is proportional to the excitation power $P$. Additionally, the intensity of the spin wave excited by the antenna is also proportional to $P$ at $P < 1$ mW, where the nonlinear damping is not active (point-up triangles in Fig. 4(a)). These facts allow us to estimate the critical precession angle corresponding to the onset of the nonlinear damping at the threshold power $P=1$ mW: $\theta \approx 9°$. At this angle, the nonlinear shift of the wavevector $dk = 0.27$ µm$^{-1}$, which corresponds to the modification of the wavelength by about 6%.

At powers $P > 1$ mW, the estimation of precession angles is less straightforward. Due to the influence of the nonlinear damping, the energy becomes transferred from the directly excited coherent spin wave into incoherent short-wavelength spin-wave modes each of them reducing the saturation magnetization accordingly. Under these conditions, the reduction of the static magnetization causing the spectral shift is determined not only by the intensity of the coherent wave but also by those of the incoherent modes. Therefore, we introduce an effective total precession angle, $\theta_{tot}$, which is generally larger than the coherent spin-wave precession angle, $\theta_{SW}$. The former can be directly found from the analysis of the shift of the dispersion spectrum (point-up triangles in Fig. 5): at the maximum power $P=3$ mW, the effective total angle $\theta_{tot} \approx 17°$. The angle $\theta_{SW}$ can be estimated based on the analysis of the power dependence of the intensity of the coherent spin wave (point-up triangles in Fig. 4(a)). We extrapolate the linear fit of the experimental data (line in Fig. 4(a)) to $P=3$ mW and find the ratio of this value to the experimentally observed intensity at this power, $\approx 1.7$. This ratio is approximately equal to $\sin^2(\theta_{tot})/\sin^2(\theta_{SW})$, where $\sin^2(\theta_{SW})$ is proportional to the measured BLS intensity while $\sin^2(\theta_{tot})$ is proportional to the microwave power. This allows us to estimate $\theta_{SW} \approx 13°$. As seen from the data of Figs. 4(a) and Fig. 5, further increase in the excitation power above 3 mW is expected to result in the further increase of $\theta_{tot}$, while $\theta_{SW}$ shows a clear tendency to saturation. We emphasize that, although the spectral shift is determined by $\theta_{tot}$, the linear increase of its value with the increase in *P* is only observed near the antenna, while at distances of several micrometers, this angle also exhibits saturation (circles and point-down triangles in Fig. 5).

We finally discuss possible effects of the heating of the sample by the intense microwave radiation. According to the theory of spin-wave interactions (Ref. 2), the increase of the temperature is not expected to noticeably affect the nonlinear damping. This was also shown experimentally in, e.g., Ref. 35. However, the heating can potentially contribute to the observed shift of the spin-wave spectrum. Since the heating results in an increase of the intensities of incoherent spin-wave modes, it contributes to the reduction of the saturation magnetization and causes an additional frequency shift. Thermally induced shift can be distinguished based on its slow temporal dynamics. Therefore, we performed additional time-resolved measurements, where the excitation was applied in the form of pulses with the duration of 5-50 µs and the spectral shift was analyzed in the time domain with the resolution down to 2 ns. The measurements at the largest microwave power of 3 mW did not reveal any slowly varying contribution suggesting that the heating effects are negligible in the studied system.

## IV. CONCLUSIONS

In conclusion, we have shown that the effect of the nonlinear spectral shift enabling the controllability of the wavelength of spin waves by their intensity is a complex phenomenon, which is strongly affected by the nonlinear spin-wave damping. This effect becomes pronounced at precession angles exceeding 9° and results in a spatially-dependent controllability of the wavelength. The efficient controllability can only be achieved at small distances from the excitation point, whereas the controllability becomes relatively poor several micrometers away. Additionally, the fast spatial decay caused by the nonlinear damping results in the strong spatial variation of the wavelength near the excitation point. These findings are critically important for the development of efficient nonlinear magnonic devices utilizing the effect of the nonlinear spectral shift.

ACKNOWLEDGMENTS

This work was supported in part by the Deutsche Forschungsgemeinschaft (DFG, German Research Foundation) – Project-ID 433682494 – SFB 1459 and TRR227 TP B02.

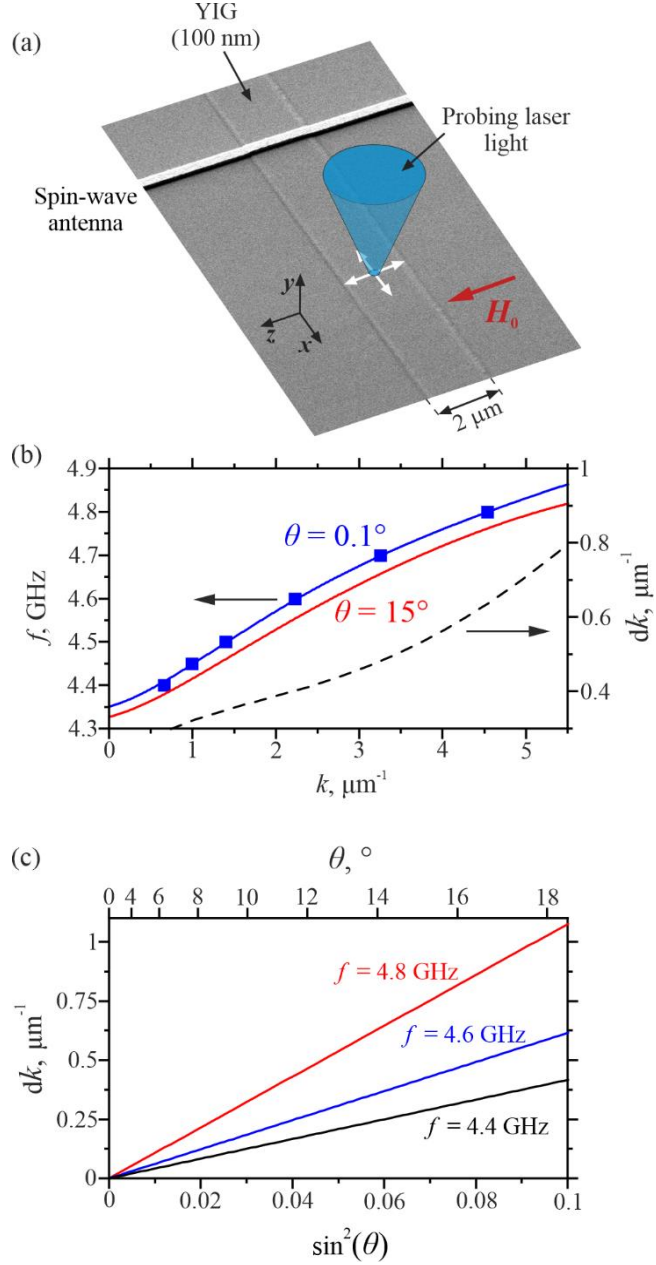

FIG. 1. (a) Schematics of the experiment. (b) Solid curves – calculated dispersion curves of spin waves in the YIG waveguide corresponding to different angles of magnetization precession, as labeled. Dashed curve – variation of the wavenumber caused by the increase in the precession angle from 0.1 to 15°. Symbols – dispersion curve measured by BLS at low excitation power $P = 0.1$ mW. (c) Dependence of the nonlinear variation of the wavenumber on the precession angle calculated at different frequencies, as labeled. The data are obtained at $H_0=1000$ Oe.

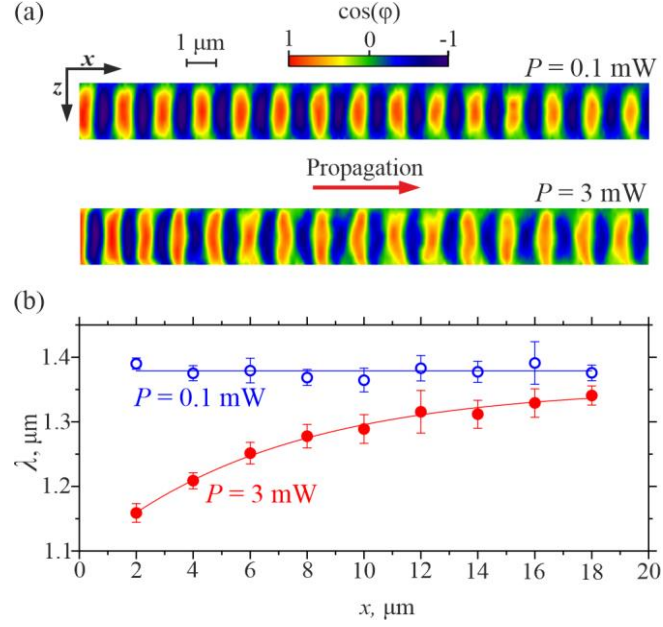

FIG. 2. (a) Representative spin-wave phase maps recorded by BLS at $f = 4.8$ GHz and $P = 0.1$ and 3 mW, as labeled. (b) Spatial dependence of the wavelength of spin waves obtained from the Fourier analysis of the phase maps. Symbols – experimental data. Curves – guide for the eye. The data are obtained at $H_0=1000$ Oe.

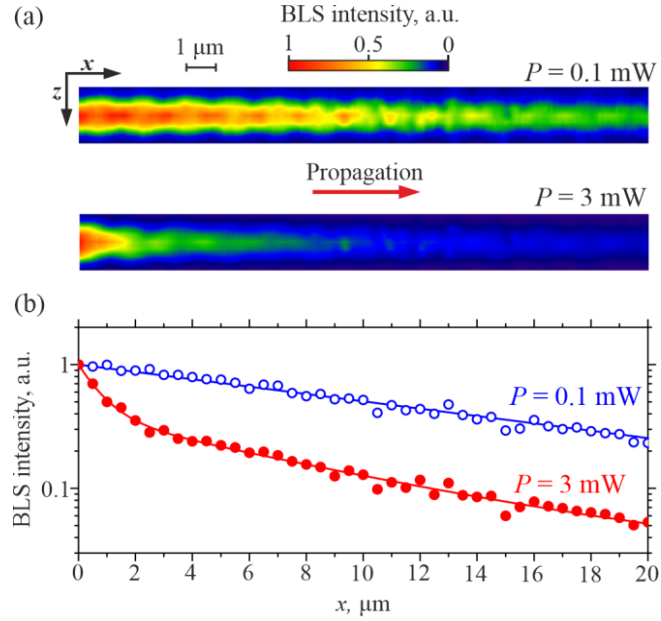

FIG. 3. (a) Spin-wave intensity maps recorded by BLS at $f = 4.8$ GHz and $P = 0.1$ and 3 mW, as labeled. (b) Normalized spatial dependence of the BLS intensity integrated over the waveguide width. Note the logarithmic scale of the vertical axis. Symbols – experimental data. Curves – exponential fit of the data obtained at $P = 0.1$ mW and double-exponential fit of the data obtained at $P = 3$ mW. The data are obtained at $H_0=1000$ Oe.

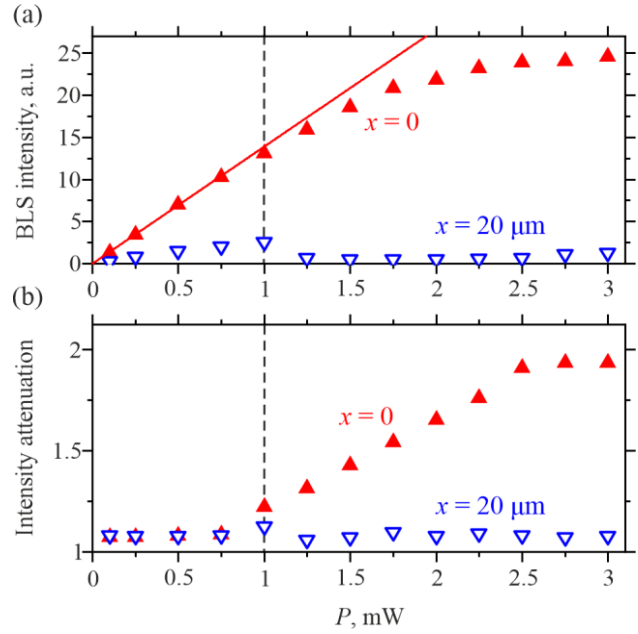

FIG. 4. (a) Power dependence of the BLS intensity recorded at $x = 0$ and 20 μm, as labeled. Line – linear fit of the experimental data at $P < 1$ mW. (b) Power dependence of the factor, by which the spin-wave intensity is attenuated over a propagation distance of 1 μm, at $x = 0$ and 20 μm, as labeled. Vertical dashed line in (a) and (b) marks the threshold power, at which the nonlinear damping emerges. The data are obtained at $H_0 = 1000$ Oe.

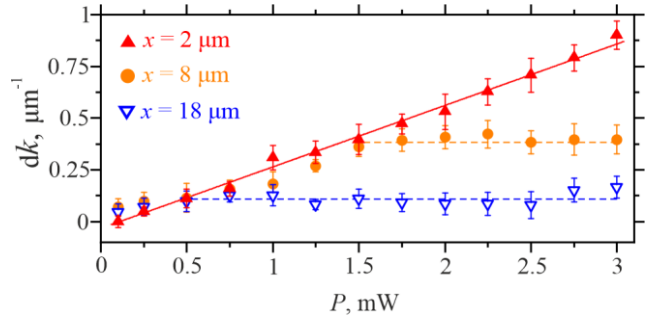

FIG. 5. Power dependences of the nonlinear variation of the wavenumber of spin waves at different distances from the antenna, as labeled. Symbols – experimental data. Solid line – linear fit of the data obtained at $x = 2$ μm. Horizontal dashed lines mark the saturation values at $x = 8$ and 18 μm. The data are obtained at $H_0$=1000 Oe.